\documentclass{amsart}
\usepackage{graphicx}
\theoremstyle{definition}

\theoremstyle{remark}

\numberwithin{equation}{section}
\RequirePackage{amssymb}
\usepackage{amssymb}
\usepackage{latexsym}
\usepackage{fancyhdr}
\usepackage[english]{babel}
\usepackage[utf8]{inputenc}
\usepackage[T1]{fontenc}
\usepackage{lmodern}
\usepackage{float}

\RequirePackage{ifpdf}

 \address{ Instituto de Ciencias Nucleares, Universidad Nacional
Aut\'onoma de M\'exico, Apartado Postal 70-543, M\'exico 04510 DF,
M\'exico.}
\email{carlos.margalli@nucleares.unam.mx}
\email{vergara@nucleares.unam.mx}
\date{January 10, 2014}
\author{Carlos A. Margalli \and J. David Vergara}

\title{Quantization of a Complex Higher Order Derivative Theory using Path Integrals}
\begin{document}
\maketitle

\begin{abstract}
This work addresses the quantization of a self-interacting higher
order time derivative theory using path integrals. To quantize this
system and avoid the problems of energy not bounded from below and
states of negative norm, we observe the following steps: 1) We
extend the theory to the complex plane and in this sense we double
the degrees of freedom. 2) We add a total derivative to fix the
convenient boundary conditions. 3) We check that the complex
structure is consistent. 4) To map from the complex space to the
real space we introduce reality conditions as second class
constraints and we check that the interactions do not generate more
constraints. 5) We built the measure of the complex path integral
and we show that including currents the theory is projected to a
self-interacting real theory that is renormalizable.
\end{abstract}

\section{Introduction}
The aim of this paper is to study the quantization of interacting
high order time derivative theories using path integrals
\cite{Pai,berna,Gro,Eli,Cer,Tai,Nakamura,Andr}. It is common
knowledge that the quantization of these kind of systems cannot be
done consistently if we add interactions, since the decouple ghosts
reappear again in this instance \cite{Antoniadis}. Then, one is
forced to select in these cases between a Hamiltonian no bounded
from below, {\it i.e.} no vacuum state, or states with negative
norm, {\it i.e.} ghosts \cite{Woodard, smilga}. So it seems that
there is no way to quantize consistently these systems. However, in
recent years there are a renewal interest in these kind of theories,
because they are an example of systems with Lorentz symmetry
breaking \cite{pospe, koste}. The usual way to treat these theories
is to consider a perturbative approach \cite{Eli, Tai} or an
effective approach \cite{berna, Gro}. Both approaches have the
disadvantage that in the procedure they lost degrees of freedom,
{\it i.e.} at the end we only have a description of the low energy
modes. If, we consider that these theories are only the effective
part of a more fundamental theory, this kind of approach have a full
justification. However, if we consider that these theories are
fundamental we are losing physically relevant information of the
theory. An alternative approach was proposed in \cite{Berg} in order
to avoid loosing physically relevant information. However, this
approach uses the gauge/gravity correspondence then until now has
only been formulated in the case of Anti-de-Sitter space.

To capture this lost information in \cite{dec}, was proposed a
procedure to quantize {\it complex high order derivative theories}
at the level of particles and in \cite{Margalli} this work was
extended to field theory using a canonical quantization. In this
paper we generalize this procedure using path integrals and we
analyze the full consistency of the procedure. The key point of our
method is not quantize directly the real high order time derivative
theory. Instead of that we quantize a complexified version of the
theory and we project this theory to the real space. To realize this
projection we analyze the consistency of the problem in the context
of Dirac's quantization of systems with constraints and we show what
type of interactions can be introduced consistently in the model.
The first step in the procedure is to complexify the original high
order derivative theory. This implies that the original degrees of
freedom of the theory are duplicated. The second step is to add a
total derivative to the action, this derivative change the variables
that are fixed at the boundary and consequence selects what fields
are associated to the real degrees of freedom, this two steps are
analyzed in section \ref{analogo}. In Section 3, we will look more
closely at the complex structure of the theory and we check that the
complexified equations of motion satisfy the Cauchy-Riemann
conditions. The next step is to map the complex higher order
derivative theory to a theory of real scalar fields. To make this
mapping consistently we introduce a set of reality conditions that
reduce the degrees of freedom of the complex theory and turn out the
new theory real. We show that this reality conditions can be
interpreted as second class constraints and we show that the free
theory is closed under the evolution of this constraints. In Section
5 we introduce the interactions and we show that there are several
types of interactions that are consistent with the reality
conditions in the sense that the addition of these terms to the
total Hamiltonian do not produce additional constraints, and in this
sense we prove that our procedure is fully consistent inclusive when
we add interactions. Section 6 is devoted to the quantization of the
theory using path integrals. We introduce a measure in the extended
space that includes the full set of second class constraints. This
measure will project the complex theory to a real one. Because we
add sources, to properly take into account the interactions. We add
as additional conditions the relationships that we obtain for the
sources from the equations of motion and reality conditions. To
apply our results in Section 7 we quantize a high order derivative
theory that corresponds to the Schwinger model via bosonization
\cite{schwinger}.

\section{Structure of a Complex Higher Order Derivative Theory \label{analogo}}
The Bernard-Duncan model is the most basic higher order time
derivative field theory with action given by
\begin{equation}\label{s}
S_{0}=\int\!d^{4}x\frac{1}{2}[-(\Box\varphi)^{2}+(m_{1}^{2}+m_{2}^{2})\partial_{\mu}
\varphi\partial^{\mu}\varphi-m_{1}^{2}m_{2}^{2}\varphi^{2}]
\end{equation}
where the scalar field $\varphi$ is real. In this point one can
think of quantizing this model following the usual procedure in
quantum field theory, but we will find problems as the existence of
negative norm states, the energy is unbounded from below and the
dispersion matrix is non unitary. However, these troubles can be
considered as interpretation failure following the ideas of Pais and
Uhlenbeck \cite{Pai} and the conclusions of Hawking and Hertog
\cite{Cer}, taking into account to two independent Hilbert spaces.
In spite of all, it is not possible to include interactions in the
system.

In this work we want to take a step forward and analyze an extension
to the complex plane of the Bernard-Duncan model. This extension
implies that the theory is not Hermitian. However, we will show that
this complex model can be consistently restricted to a real phase
space. This restriction is implemented using second class
constraints following the Dirac's formalism \cite{Dir}, and we show
that the constraint surface is preserved by the inclusion of some
kind of interactions.

Firstly we establish a complexification of the Bernard-Duncan theory, i.e., we define that the higher order field is complex
\begin{equation}\label{complex}
 \phi\equiv\phi_{R}+i\phi_{I}.
\end{equation}
With this complexification, the number of degrees of freedom is
duplicated. Secondly we attach a total derivative term to the
complex Lagrangian density, this does not modify the equations of
motion, but it allows to pick out the boundary conditions in terms
of the field $\phi$  and the acceleration  $\ddot{\phi}$.

The Lagrangian density attaching the total derivative term is
\begin{equation}\label{a}
S=\int\!d^{4}x\left\lbrace \frac{1}{2}[(\Box\phi)^{2}+(m_{1}^{2}+m_{2}^{2})\partial_{\mu}
\phi\partial^{\mu}\phi-m_{1}^{2}m_{2}^{2}\phi^{2}]+\partial_{\mu}\phi(\partial^{\mu}\Box\phi)\right\rbrace
\end{equation}
with a complex field $\phi$ and a Lagrangian density which is an complex analytic function.

The above (\ref{a}) is used to obtain directly the respective momenta
\begin{eqnarray}\label{momen}
 \pi_{0}=\frac{\partial\mathcal{L}}{\partial\dot{\phi}}-\partial_{\mu}\frac{\partial\mathcal{L}}{\partial\partial_{\mu}
\dot{\phi}}+\partial_{\nu}\partial_{\sigma}\frac{\partial\mathcal{L}}{\partial\partial_{\nu}\partial_{\sigma}\dot{\phi}}\\
\pi_{1}=\frac{\partial\mathcal{L}}{\partial\ddot{\phi}}-\partial_{\mu}\frac{\partial\mathcal{L}}{\partial\partial_{\mu}
\ddot{\phi}}\\
\label{momenu}
\pi_{2}=\frac{\partial\mathcal{L}}{\partial\phi^{(3)}}
\end{eqnarray}
resulting explicitly
\begin{eqnarray}\label{momentao}
 \pi_{0}=\phi^{(3)}-2\nabla^{2}\dot{\phi}+(m_{1}^{2}+m_{2}^{2})\dot{\phi}\\
\pi_{1}=0\\\label{momentat}
\pi_{2}=\dot{\phi}
\end{eqnarray}
which are separable in real and imaginary parts. From these momenta we obtain four constraints which later we will
study using  Dirac's theory of constraints \cite{Dir,Hen}.

To solve the equations of motion of (\ref{a}) we need to specify the
initial conditions $\phi(\vec{x},t=t_{0})$, $\dot{\phi}(\vec{x},
t=t_{0})$, $\ddot{\phi}(\vec{x},t=t_{0})$ and
$\phi^{(3)}(\vec{x},t=t_{0})$. The above define the configuration
space of the Lagrangian theory and it shows that we  have 8 linearly
independent solutions, because our theory is complex.

From the Ostrogradsky Hamiltonian description of the third order
theory (\ref{a}) \cite{os}, we define the independent fields
\begin{equation}\label{fields}
\phi=\phi,\qquad\eta=\dot{\phi},\qquad\xi=\ddot{\phi},
\end{equation}
that will be used to establish clearly the appearance of
constraints.

Fields and momenta allow us to introduce the Hamiltonian theory that
is complex, but it is not clear the way we introduce higher order
quantities. In next subsection, by means of Schwinger action's
principle we argue reasons to introduce these fields (\ref{fields})
and these momenta (\ref{momen}) that result in a consistent
Hamiltonian theory.

\subsection{The Schwinger Variational Principle}

In order to show the effect of boundary conditions, we use the
Schwinger variational principle \cite{Ditt}, applied to the action
(\ref{a}) resulting
\begin{equation}
 \delta S=\int\! d^{4}x -\frac{1}{2}\phi\left\{ \Box(\Box)+(m_{1}^{2}+m_{2}^{2})\Box+m_{1}^{2}m_{2}^{2}\right\}\phi
+\int \! d^{3}x\left\{\pi_{0}\delta\phi+\pi_{2}\delta\xi\right\},
\end{equation}
where we find that the fields $\phi$ and $\xi$ are naturally fixed
on the boundary and in this sense the definitions in (\ref{momen})
-(\ref{momenu}) are consistent. Also, from the variation we see that
there are only 2 complex degrees of freedom in the configuration
space, 4 in the phase space $(\phi, \xi, \pi_{0}, \pi_{2})$, instead
of the 3 complex that we see in (\ref{fields}), and this implies
that our theory has constraints.

The Hamiltonian density resulting from Ostrogradsky's Theory is
\begin{eqnarray}\label{ostrogradsky}
 \mathcal{H}=\pi_{1}\xi+\pi_{0}\pi_{2}+2\pi_{2}\nabla^{2}\eta-(m_{1}^{2}+m_{2}^{2})\eta\pi_{2}
-\frac{1}{2}\xi^{2}-\frac{1}{2}(\nabla^{2}\phi)^{2}+\xi\nabla^{2}\phi\\\nonumber
-\frac{(m_{1}^{2}+m_{2}^{2})}{2}\eta^{2}+\frac{(m_{1}^{2}+m_{2}^{2})}{2}\nabla\phi\cdotp\nabla\phi
+\frac{m_{1}^{2}m_{2}^{2}}{2}\phi^{2}
-\eta\nabla^{2}\eta+(m_{1}^{2}+m_{2}^{2})\eta^{2}\\ \nonumber
-\xi\nabla^{2}\phi+(\nabla^{2}\phi)^{2}+\pi_{0}\pi_{2}-\eta\pi_{0}.
\end{eqnarray}

The Hamiltonian density (\ref{ostrogradsky}) in terms of real and imaginary parts is
\begin{eqnarray}\label{hamrealim}
 \mathcal{H}_{R}=\pi_{1R}\xi_{R}+\pi_{0R}\pi_{2R}+2\pi_{2R}\nabla^{2}\eta_{R}-(m_{1}^{2}+m_{2}^{2})\eta_{R}\pi_{2R}
-\frac{1}{2}\xi_{R}^{2}\\\nonumber
-\frac{1}{2}(\nabla^{2}\phi_{R})^{2}+\xi_{R}\nabla^{2}\phi_{R}-\frac{(m_{1}^{2}+m_{2}^{2})}{2}\eta_{R}^{2}+\frac{(m_{1}^{2}
+m_{2}^{2})}{2}\nabla\phi_{R}\cdotp\nabla\phi_{R}
\\ \nonumber
+\frac{m_{1}^{2}m_{2}^{2}}{2}\phi_{R}^{2}-\eta_{R}\nabla^{2}\eta_{R}+(m_{1}^{2}+m_{2}^{2})\eta_{R}^{2}
-\xi_{R}\nabla^{2}\phi_{R}+(\nabla^{2}\phi_{R})^{2}\\ \nonumber
+\pi_{0R}\pi_{2R}-\eta_{R}\pi_{0R}\\\nonumber
-[\pi_{1I}\xi_{I}+\pi_{0I}\pi_{2I}+2\pi_{2I}\nabla^{2}\eta_{I}-(m_{1}^{2}+m_{2}^{2})\eta_{I}\pi_{2I}
-\frac{1}{2}\xi_{I}^{2}\\\nonumber
-\frac{1}{2}(\nabla^{2}\phi_{I})^{2}+\xi_{I}\nabla^{2}\phi_{I}-\frac{(m_{1}^{2}+m_{2}^{2})}{2}\eta_{I}^{2}+\frac{(m_{1}^{2}
+m_{2}^{2})}{2}\nabla\phi_{I}\cdotp\nabla\phi_{I}
\\ \nonumber
+\frac{m_{1}^{2}m_{2}^{2}}{2}\phi_{I}^{2}-\eta_{I}\nabla^{2}\eta_{I}+(m_{1}^{2}+m_{2}^{2})\eta_{I}^{2}
-\xi_{I}\nabla^{2}\phi_{I}+(\nabla^{2}\phi_{I})^{2}\\\nonumber
+\pi_{0I}\pi_{2I}-\eta_{I}\pi_{0I}],\\ \nonumber
\mathcal{H}_{I}=[\pi_{1R}\xi_{I}+\pi_{1I}\xi_{R}+\pi_{0R}\pi_{2I}+\pi_{0I}\pi_{2R}+2\pi_{2R}\nabla^{2}\eta_{I}
+2\pi_{2I}\nabla^{2}\eta_{R}\\\label{hamrealimdos}
-(m_{1}^{2}+m_{2}^{2})\eta_{R}\pi_{2I}-(m_{1}^{2}+m_{2}^{2})\eta_{I}\pi_{2R}-\xi_{R}\xi_{I}
-\nabla^{2}\phi_{R}\nabla^{2}\phi_{I}\\\nonumber
+\xi_{R}\nabla^{2}\phi_{I}+\xi_{I}\nabla^{2}\phi_{R}-(m_{1}^{2}+m_{2}^{2})\eta_{R}\eta_{I}
+(m_{1}^{2}+m_{2}^{2})\nabla\phi_{R}\cdotp\nabla\phi_{I}\\\nonumber
+m_{1}^{2}m_{2}^{2}\phi_{R}\phi_{I}-\eta_{R}\nabla^{2}\eta_{I}-\eta_{I}\nabla^{2}\eta_{R}
+2(m_{1}^{2}+m_{2}^{2})\eta_{R}\eta_{I}\\\nonumber
-\xi_{R}\nabla^{2}\phi_{I}-\xi_{I}\nabla^{2}\phi_{R}+2(\nabla^{2}\phi_{R})(\nabla^{2}\phi_{I})\\\nonumber
+\pi_{0R}\pi_{2I}+\pi_{0I}\pi_{2R}-\eta_{R}\pi_{0I}-\eta_{I}\pi_{0R}].\\\nonumber
\end{eqnarray}
In this part we have found a real phase space with 12 real degrees
of freedom, but with the Schwinger variational method we have 8
degrees of freedom. In fact, it suggests which the Hamiltonian
theory is incomplete and we need to incorporate the constraints and
in the next section this problem is faced using the Dirac's Method.
Furthermore, in the spirit of the Complex Hamiltonian in the next
section we analyze how to introduce a complex symplectic structure
in such way that the Hamiltonian equations satisfy identically the
Cauchy-Riemann conditions and in this way the evolution of the
theory be analytical.

\section{Hamilton's Equations and Cauchy-Riemann equations}
In this section we study the complex Hamiltonian structure and we
select the correct symplectic structure in such way the classical
evolution respects the analyticity of the system.

\subsection{Complex Structure}

In this part of the work, we are going to determine the minimal
element that produces the temporal evolution in this complex
description. From the separation in real and complex parts
(\ref{complex}), we can establish a deeper analysis, in this case
the Legendre transformation for the Hamiltonian density
(\ref{ostrogradsky}) is
\begin{eqnarray}\label{denlan}
 \mathcal{L}=\dot{\phi}_{R}\pi_{0R}-\dot{\phi}_{I}\pi_{0I}+\dot{\eta}_{R}\pi_{1R}-\dot{\eta}_{I}\pi_{1I}
+\dot{\xi}_{R}\pi_{2R}-\dot{\xi}_{I}\pi_{2I}-\mathcal{H}_{R}\\\nonumber
+i[\dot{\phi}_{I}\pi_{0R}+\dot{\phi}_{R}\pi_{0I}+\dot{\eta}_{I}\pi_{1R}+\dot{\eta}_{R}\pi_{1I}
+\dot{\xi}_{R}\pi_{2I}+\dot{\xi}_{I}\pi_{2R}-\mathcal{H}_{I}].
\end{eqnarray}
where we introduce two symplectic structures. One for the real part
of the Hamiltonian density and another one for the imaginary part.

To begin with we define a new notation that establish a more compact
description
\begin{eqnarray}
 \Theta^{a}_{A}=(\phi_{R},\eta_{R},\xi_{R},\phi_{I},\eta_{I},\xi_{I}),\label{aaa}\\
\Pi_{bB}=(\pi_{0R},\pi_{1R},\pi_{2R},\pi_{0I},\pi_{1I},\pi_{2I})\label{bbb},
\end{eqnarray}
with the index $a$ running from $a=(\phi,\eta,\xi)$ and
$b=(\pi_{0},\pi_{1},\pi_{2})$ and the subscripts $A,B=(R,I)$ run
over the real and imaginary parts. From the variation of the
Lagrangian density (\ref{denlan}) we obtain
\begin{eqnarray}
  \dot{\Theta}^{a}_{R}=\frac{\partial \mathcal{H}_{R}}{\partial \Pi_{aR}}=\frac{\partial \mathcal{H}_{I}}{\partial \Pi_{aI}},
  \qquad \dot{\Theta}^{a}_{I}=\frac{\partial \mathcal{H}_{I}}{\partial \Pi_{aR}}=-\frac{\partial
  \mathcal{H}_{R}}{\partial \Pi_{aI}},\\
 \dot{\Pi}_{aR}=\frac{\partial \mathcal{H}_{R}}{\partial \Theta^{a}_{R}}=\frac{\partial
 \mathcal{H}_{I}}{\partial \Theta^{a}_{I}},\qquad \dot{\Pi}_{aI}=-\frac{\partial \mathcal{H}_{I}}{\partial
 \Theta^{a}_{R}}=\frac{\partial \mathcal{H}_{R}}{\partial
 \Theta^{a}_{I}}.
\end{eqnarray}
This is the full set of Hamilton equations and this system satisfies
the Cauchy-Riemann conditions and in consequence the evolution given
by these equations is analytical. From the symplectic structure
given in (\ref{denlan}) the Poisson brackets are
\begin{equation}\label{masmenos}
\{\Theta^{a}_{A}(t,\vec{x}),\Pi_{bB}(t,\vec{x'})\}=\mathcal{J}_{AB}\delta_{b}^{ a}\delta^{3}(\vec{x}-\vec{x'}),
\end{equation}
where we have defined $\mathcal{J}_{AB}$ as
\begin{displaymath}
\mathcal{J}_{AB}=\left\{ %
\begin{array}{cc}
 1 & \textrm{si}\ A=B=R\\
0  & \textrm{si}\ \ A\neq B\\
-1 & \textrm{si}\ A=B=I
\end{array}\right.
\end{displaymath}
From this expression we obtain the general definition
\begin{eqnarray}\label{brac}
\{ F,G \}=\int \! d^{3}x'(\frac{\delta F}{\delta \phi_{R}}\frac{\delta G}{\delta \pi_{0R}}
-\frac{\delta F}{\delta \pi_{0R}}\frac{\delta G}{\delta \phi_{R}})-(\frac{\delta F}{\delta \phi_{I}}\frac{\delta G}
{\delta \pi_{0I}}
-\frac{\delta F}{\delta \pi_{0I}}\frac{\delta G}{\delta \phi_{I}})\\\nonumber
+(\frac{\delta F}{\delta \eta_{R}}\frac{\delta G}{\delta \pi_{1R}}
-\frac{\delta F}{\delta \pi_{1R}}\frac{\delta G}{\delta \eta_{R}})-(\frac{\delta F}{\delta \eta_{I}}\frac{\delta G}
{\delta \pi_{1I}}
-\frac{\delta F}{\delta \pi_{1I}}\frac{\delta G}{\delta \eta_{I}})\\\nonumber
+(\frac{\delta F}{\delta \xi_{R}}\frac{\delta G}{\delta \pi_{2R}}
-\frac{\delta F}{\delta \pi_{2R}}\frac{\delta G}{\delta \xi_{R}})-(\frac{\delta F}{\delta \xi_{I}}\frac{\delta G}
{\delta \pi_{2I}}
-\frac{\delta F}{\delta \pi_{2I}}\frac{\delta G}{\delta \xi_{I}}).\\\nonumber
\end{eqnarray}
It is important to mention that in the parenthesis(\ref{masmenos})
there are several terms with oppositive signs, they appear as a
natural consequence of the complex structure and Legendre
transformation (\ref{denlan}).

\subsection{Constraints in the System} In order to analyze the
dynamics of the system we will use the Poisson brackets of
(\ref{brac}) and now
 we consider the Dirac's Theory of constraints in order to
handle systematically the restrictions which we have found in the definition of the momenta.

From the momenta (\ref{momentao})-(\ref{momentat}) we get four
primary constraint since we have divide the real and imaginary parts
\begin{eqnarray}\label{cosmouno}
 \gamma_{1}=\pi_{1R}, \qquad \gamma_{2}=\pi_{1I},\\
\gamma_{3}=\pi_{2R}-\eta_{R},\qquad \gamma_{4}=\pi_{2I}-\eta_{I}.\nonumber
\end{eqnarray}
These constraints satisfy
\begin{equation}\label{gamma}
 \{ \gamma_{a} ,\gamma_{b} \}= \mathcal{C}_{ab}\delta(\vec{x}-\vec{x'})
\end{equation}
and
\begin{equation}\label{matrix}
\mathcal{C}_{ab}=\left(
\begin{array}{rccl}
0 & 0 & 1 & 0\\
0 & 0 & 0 & -1\\
-1 & 0 & 0 & 0\\
0 & 1 & 0 & 0
\end{array}
\right),
\end{equation}
with determinant given by
\begin{equation}
\det \left( \mathcal{C}_{ab}\right) =1.
\end{equation}
The temporal evolution of constraints is accomplished through the
Cauchy-Riemann equations and we find that by the analyticity, it is
associated with either the real or imaginary parts of the complex
Hamiltonian density
\begin{eqnarray}\label{uno}
 \dot{\gamma}_{1}=\int\!d^{3}x'\{\gamma_{1},\mathcal{H}_{R}+\alpha_{1}\gamma_{1}+\alpha_{2}\gamma_{2}+\alpha_{3}\gamma_{3}
+\alpha_{4}\gamma_{4}\}\\\nonumber
=[-2\nabla^{2}+(m_{1}^{2}+m_{2}^{2})]\gamma_{3}+\pi_{0R}+\alpha_{3}\approx0, \\
\dot{\gamma}_{2}=[-2\nabla^{2}+(m_{1}^{2}+m_{2}^{2})]\gamma_{4}+\pi_{0I}-\alpha_{4}\approx0,\\
\dot{\gamma}_{3}=-\gamma_{1}-\nabla^{2}\phi_{R}-\alpha_{1}\approx0,\\\label{dos}
\dot{\gamma}_{4}=-\gamma_{2}-\nabla^{2}\phi_{I}+\alpha_{2}\approx0,
\end{eqnarray}
where we see that these constraints form a complete set, since from
(\ref{uno})-(\ref{dos}) we obtain the Lagrange multipliers
$\alpha$'s and because the expression (\ref{gamma}) this set is made
of second class constraints. This of  course implies that we pass
from 12  to 8 real degrees of freedom of the complex phase space.
This constraint (\ref{cosmouno}) define a new symplectic structure
through the Dirac's brackets
\begin{eqnarray}\label{padedi}
 \{ F(t,\vec{x}_{0}),G(t,\vec{x}) \}^{*}=\{ F(t,\vec{x}_{0}),G(t,\vec{x}) \}\label{diracos}\\\nonumber
-\int\!dx'dx''\{F(t,\vec{x}_{0}),\gamma_{a}(t,\vec{x'}) \}C^{ab}\delta(\vec{x'}-\vec{x''})
\{\gamma_{b}(t,\vec{x''}),G(t,\vec{x})\}.
\end{eqnarray}
From the definition (\ref{diracos}) the new symplectic structure  results
\begin{eqnarray}\label{dirbra}
 \{ \phi_{R}(t,\vec{x}),\pi_{0R}(t,\vec{x}_{0})\}^{*}=\delta^{3}(\vec{x}-\vec{x}_{0})\\
\nonumber
 \{ \phi_{I}(t,\vec{x}),\pi_{0I}(t,\vec{x}_{0})\}^{*}=-\delta^{3}(\vec{x}-\vec{x}_{0})\\
\nonumber
\{ \eta_{R}(t,\vec{x}),\xi_{R}(t,\vec{x}_{0})\}^{*}=-\delta^{3}(\vec{x}-\vec{x}_{0})\\\nonumber
 \{ \eta_{I}(t,\vec{x}),\xi_{I}(t,\vec{x}_{0})\}^{*}=\delta^{3}(\vec{x}-\vec{x}_{0})\\
\nonumber
\{ \xi_{R}(t,\vec{x}),\pi_{2R}(t,\vec{x}_{0})\}^{*}=\delta^{3}(\vec{x}-\vec{x}_{0})\\
\nonumber
 \{ \xi_{I}(t,\vec{x}),\pi_{I}(t,\vec{x}_{0})\}^{*}=-\delta^{3}(\vec{x}-\vec{x}_{0}).\\
\nonumber
\end{eqnarray}
We will use this new  symplectic structure for the following
computations, and in this way we have incorporated the correct
boundary conditions.

To quantize the system  we need to promote the Dirac's brackets
(\ref{dirbra}) to commutators and this procedure is quite simple
since the matrix (\ref{matrix}) is constant, then we can use the
constraints (\ref{cosmouno}) directly in the Hamiltonian and from
them eliminate the variables $(\eta_{R}, \eta_{I}, \pi_{1R},
\pi_{1I})$.

In the reduced space the Hamiltonian density is given by
\begin{eqnarray}
\mathcal{H}_{C}=\mathcal{H}_{CR}+i\mathcal{H}_{CI}\label{hamreaev}\\
\mathcal{H}_{CR}=\pi_{0R}\pi_{2R}-\frac{1}{2}\xi^{2}_{R}-\frac{(m_{1}^{2}+m_{2}^{2})}{2}\pi_{2R}^{2}
+\frac{m_{1}^{2}m_{2}^{2}}{2}\phi_{R}^{2}+\frac{1}{2}(\nabla^{2}\phi_{R})^{2}\\\nonumber
+\frac{(m_{1}^{2}+m_{2}^{2})}{2}(\nabla\phi_{R})^{2}+\pi_{2R}\nabla^{2}\pi_{2R}\\\nonumber
-[\pi_{0I}\pi_{2I}-\frac{1}{2}\xi^{2}_{I}
-\frac{(m_{1}^{2}+m_{2}^{2})}{2}\pi_{2I}^{2}+\frac{m_{1}^{2}m_{2}^{2}}{2}\phi_{I}^{2}+\frac{1}{2}(\nabla^{2}\phi_{I})^{2}\\
\nonumber
+\frac{(m_{1}^{2}+m_{2}^{2})}{2}(\nabla\phi_{I})^{2}+\pi_{2I}\nabla^{2}\pi_{2I}]
\\
\mathcal{H}_{CI}=
[\pi_{0R}\pi_{2I}+\pi_{0I}\pi_{2R}-\xi_{R}\xi_{I}-(m_{1}^{2}+m_{2}^{2})\pi_{2R}\pi_{2I}\\\nonumber
+m_{1}^{2}m_{2}^{2}\phi_{R}\phi_{I}+\nabla^{2}\phi_{R}\nabla^{2}\phi_{I}\\\nonumber
+(m_{1}^{2}+m_{2}^{2})\nabla\phi_{R}\cdotp\nabla\phi_{I}+\pi_{2R}\nabla^{2}\pi_{2I}+\pi_{2I}\nabla^{2}\pi_{2R}].
\end{eqnarray}
In this way at this moment the dynamics of the system is given by the Hamiltonian density (\ref{hamreaev}) with
the symplectic structure (\ref{dirbra}).

The reduced Hamiltonian density written in terms of complex
variables $(\phi,\xi,\pi_{0}, \pi_{2})$ is
\begin{eqnarray}\label{hcanonica}
\mathcal{H}_{C}=\pi_{0}\pi_{2}-\frac{1}{2}\xi^{2}-\frac{(m_{1}^{2}+m_{2}^{2})}{2}\pi_{2}^{2}
+\frac{m_{1}^{2}m_{2}^{2}}{2}\phi^{2}+\frac{1}{2}(\nabla^{2}\phi)^{2}\\\nonumber
+\frac{(m_{1}^{2}+m_{2}^{2})}{2}(\nabla\phi)^{2}+\pi_{2}\nabla^{2}\pi_{2}.\nonumber
\end{eqnarray}
The Hamiltonian density (\ref{hcanonica}) is tightly related to the
Bernard-Duncan Hamiltonian density \cite{dec, berna}. We can see it
by means of the canonical transformation
\begin{eqnarray}\label{tiktok}
 \phi(\vec{x},t)=\phi(\vec{x},t),\qquad \pi_{\phi}(\vec{x},t)=\pi_{0}(\vec{x},t)+\triangledown^{2}\pi_{2}(\vec{x},t),\\
\pi_{\dot{\phi}}=-\xi+\nabla^{2}\phi(\vec{x},t),\qquad \dot{\phi}(\vec{x},t)=\pi_{2}(\vec{x},t),\nonumber
\end{eqnarray}
that imply
\begin{eqnarray}\label{cons}
\mathcal{H}_{D}=\pi_{\phi}\dot{\phi}-\frac{1}{2}\pi_{\dot{\phi}}^{2}-\frac{(m_{1}^{2}+m_{2}^{2})}{2}\dot{\phi}^{2}
+\frac{(m_{1}^{2}+m_{2}^{2})}{2}(\triangledown\phi)^{2}+\frac{m_{1}^{2}m_{2}^{2}}{2}\phi^{2}\\
-\triangledown\phi\cdotp\triangledown\pi_{\dot{\phi}},\nonumber
\end{eqnarray}
where it should be taken into account that our theory is complex.

Now, it is possible to apply the four constraints (\ref{cosmouno})
to the Lagrangian density (\ref{denlan}) which confirm the Dirac
brackets
\begin{eqnarray}\label{lagranca}
 \mathcal{L}_{C}=\pi_{0}\dot{\phi}+\pi_{2}\dot{\xi}-\mathcal{H}_{C}=\pi_{0R}\eta_{R}-\pi_{0I}\eta_{I}
+\pi_{2R}\dot{\xi}_{R}-\pi_{2I}\dot{\xi}_{I}-\mathcal{H}_{CR}\\ \nonumber
+i[\pi_{0R}\eta_{I}+\pi_{0I}\eta_{R}+\pi_{2R}\dot{\xi}_{I}+\pi_{2I}\dot{\xi}_{R}-\mathcal{H}_{CI}],
\end{eqnarray}
and as we proceeded it is necessary to set a new compact notation
\begin{eqnarray}
 \Lambda^{c}_{C}=(\phi_{R},\xi_{R},\phi_{I},\xi_{I}),\\
\Upsilon_{dD}=(\pi_{0R},\pi_{2R},\pi_{0I},\pi_{2I}),
\end{eqnarray}
with $c=\phi, \xi$ for the superscript of $\Lambda$, $d=\pi_{0},
\pi_{2}$ for the subscript of $\Upsilon$ and $C,D=R,I$. From the
variation of the Lagrangian  density (\ref{lagranca}) result the
Cauchy-Riemann and Hamilton equations
\begin{eqnarray}
  \dot{\Lambda}^{a}_{R}=\frac{\partial \mathcal{H}_{CR}}{\partial \Upsilon_{aR}}=\frac{\partial \mathcal{H}_{CI}}{\partial \Upsilon_{aI}},\qquad \dot{\Lambda}^{a}_{I}=\frac{\partial \mathcal{H}_{CI}}{\partial \Upsilon_{aR}}=-\frac{\partial \mathcal{H}_{CR}}{\partial \Upsilon_{aI}},\\
 \dot{\Upsilon}_{aR}=\frac{\partial \mathcal{H}_{CR}}{\partial \Lambda^{a}_{R}}=
 \frac{\partial \mathcal{H}_{CI}}{\partial \Lambda^{a}_{I}},\qquad \dot{\Upsilon}_{aI}=
 -\frac{\partial \mathcal{H}_{CI}}{\partial \Lambda^{a}_{R}}=\frac{\partial \mathcal{H}_{CR}}{\partial
 \Lambda^{a}_{I}},
\end{eqnarray}
with the Dirac bracket
\begin{eqnarray}
\{ F,G \}^{*}=\int \! d^{3}x'(\frac{\delta F}{\delta \phi_{R}}\frac{\delta G}{\delta \pi_{0R}}
-\frac{\delta F}{\delta \pi_{0R}}\frac{\delta G}{\delta \phi_{R}})-(\frac{\delta F}{\delta \phi_{I}}\frac{\delta G}
{\delta \pi_{0I}}
-\frac{\delta F}{\delta \pi_{0I}}\frac{\delta G}{\delta \phi_{I}})\\\nonumber
+(\frac{\delta F}{\delta \xi_{R}}\frac{\delta G}{\delta \pi_{2R}}
-\frac{\delta F}{\delta \pi_{2R}}\frac{\delta G}{\delta \xi_{R}})-(\frac{\delta F}{\delta \xi_{I}}\frac{\delta G}
{\delta \pi_{2I}}
-\frac{\delta F}{\delta \pi_{2I}}\frac{\delta G}{\delta \xi_{I}})\\\nonumber
\end{eqnarray}
that follows directly from the Dirac bracket (\ref{padedi}).
Explicitly, the parentheses are
\begin{equation}
\{\Lambda^{a}_{A}(t,\vec{x}),\Upsilon_{bB}(t,\vec{x'})\}^{*}=\mathcal{I}_{AB}\delta_{b}^{ a}\delta^{3}(\vec{x}-\vec{x'}),
\end{equation}
where the matrix $\mathcal{I}_{AB}$ is
\begin{displaymath}
\mathcal{I}_{AB}=\left\{ %
\begin{array}{cc}
 1 & \textrm{si}\ A=B=R\\
0  & \textrm{si}\ \ A\neq B\\
-1 & \textrm{si}\ A=B=I
\end{array}\right.
\end{displaymath}

\section{Reality Conditions in the model}
At this moment our model is still complex and the idea now is to
introduce conditions that project our system to the real space. The
idea of reality conditions that lead to real theories was proposed
initially by Ashtekar in the context of general relativity
\cite{ashtekar}. The reality conditions have been important in order
to give a physical sense to a complex theory such that these
conditions are used to cancel the effect of the imaginary part
\cite{dec}. In the original formulation of Ashtekar these conditions
were implemented through the scalar product. However, in the Ref.
\cite{MUV} it was shown that these conditions can be implemented as
second class constraints. For our case this procedure is more
useful, since in the quantization of the theory the reality
conditions can be implemented directly in the path integral. Another
important point is that the full set of reality conditions is fixed
by the evolution of the system, that means that starting from a set
of constraints we evolve this set to show if the dynamics is
consistent with the reality conditions, if appear new constraints we
include these new conditions and we finish when the algebra is
closed under the evolution.

In our case, in order to reduce to a real Hamiltonian density it is
necessary to consider as starting point two constraints that
generate a complete set
\begin{equation}\label{conreuno}
\Sigma_{1}=\pi_{0I}+\nabla^{2}\pi_{2I}-m_{2}^{2}\pi_{2I}, \qquad \Sigma_{2}=\pi_{0R}+\nabla^{2}\pi_{2R}-m_{1}^{2}\pi_{2R},
\end{equation}
and their time evolution is
\begin{equation}\label{evoluno}
 \dot{\Sigma}_{1}=\int\!d^{3}x'\left\lbrace \Sigma_{1},\mathcal{H}_{CR} \right\rbrace =   (-\nabla^{2}+m_{2}^{2})\Sigma_{4}\qquad \dot{\Sigma}_{2}=(-\nabla^{2}+m_{1}^{2})\Sigma_{3}
\end{equation}
where $\Sigma_{3}$ and $\Sigma_{4}$ are given by
\begin{equation}\label{conredos}
\Sigma_{3}=-\xi_{R}+\nabla^{2}\phi_{R}-m_{2}^{2}\phi_{R}, \qquad \Sigma_{4}=-\xi_{I}+\nabla^{2}\phi_{I}-m_{1}^{2}\phi_{I}.
\end{equation}
In order to obtain the complete set we need to establish that these
are the full set of constraints and then are closed.

The time evolution of the secondary constraints is
\begin{equation}\label{evoldos}
 \dot{\Sigma}_{3}=-\Sigma_{2}\qquad
\dot{\Sigma}_{4}=-\Sigma_{1}.
\end{equation}
In this way the system is closed. The full algebra of constraints is
\begin{equation}
 \{ \Sigma_{a} ,\Sigma_{b} \}= \mathcal{D}_{ab}\delta(\vec{x}-\vec{x'})
\end{equation}
and
\begin{equation}\label{matinv}
\mathcal{D}_{ab}=\left(
\begin{array}{rccl}
0 & 0 & 0 & -(m_{1}^{2}-m_{2}^{2})\\
0 & 0 & -(m_{1}^{2}-m_{2}^{2}) & 0\\
0 & (m_{1}^{2}-m_{2}^{2}) & 0 & 0\\
(m_{1}^{2}-m_{2}^{2}) & 0 & 0 & 0
\end{array}
\right),
\end{equation}
with determinant given by
\begin{equation}
\det \left( \mathcal{D}_{ab}\right) =(m_{1}^{2}-m_{2}^{2})^{4}
\end{equation}
and the inverse matrix exist when  $m_{1}\neq m_{2}$ resulting
\begin{equation}
\mathcal{D}^{ab}=\left(
\begin{array}{rccl}
0 & 0 & 0 &\beta^{2}\\
0 & 0 & \beta^{2} & 0\\
0 & -\beta^{2} & 0 & 0\\
-\beta^{2} & 0 & 0 & 0
\end{array}
\right),
\end{equation}
with $\beta^{2}=\frac{1}{(m_{1}^{2}-m_{2}^{2})}$. In conclusion
since, we include 4 reality conditions as second constraints, in
consequence the degrees of freedom of our theory change from 8 to 4.
We must notice that the matrix (\ref{matinv}) is invertible only in
the case that $m_1\neq m_2$. This means that for equal masses the
reality conditions are not more second class constraints, and the
theory will have several sectors \cite{Rive,smilga,Fai,Andr1}. So,
in this work we only consider the case $m_1\neq m_2$.

The phase space that we have decided to use for convenience is given by
$(\phi_{R},\phi_{I},\pi_{2R},\pi_{2I})$. In order to set the new reduced theory,
we establish the Dirac bracket
\begin{eqnarray}
 \left\lbrace F_{R}(t,\vec{x}),G_{R}(t,\vec{x}_{0}) \right\rbrace^{**}\equiv\left\lbrace  F_{R}(t,\vec{x}),G_{R}(t,\vec{x}_{0})  \right\rbrace^{*}\label{dpddu} \\
- \int\!d^{3}y \left\lbrace F_{R}(t,\vec{x}) ,\Sigma_{a}(t,\vec{y})  \right\rbrace^{*} \mathcal{D}^{ab}\left\lbrace \Sigma_{b}(t,\vec{y}) , G_{R}(t,\vec{x}_{0})   \right\rbrace^{*}\nonumber
\end{eqnarray}
and the fundamental brackets are
\begin{eqnarray}
\left\lbrace  \phi_{R}(t,\vec{x}),\pi_{2R}(t,\vec{x}_{0})  \right\rbrace^{**}=\frac{1}{(m_{1}^{2}-m_{2}^{2})}
\delta^{3}(\vec{x}-\vec{x}_{0}),\\\nonumber
\left\lbrace  \phi_{I}(t,\vec{x}),\pi_{2I}(t,\vec{x}_{0}) \right\rbrace^{**}=\frac{1}{(m_{1}^{2}-m_{2}^{2})}
\delta^{3}(\vec{x}-\vec{x}_{0}),\nonumber
\end{eqnarray}
Though, in principle we can choose a different set of starting
reality conditions this election is not  arbitrary since in this
case the cancelation of the imaginary part of the Hamiltonian
density is quite clear. This statement implies that the phase space
is established by the reality conditions (\ref{conreuno}) and
(\ref{conredos}) in order to express an inverse transformation using
these conditions.

For this reason we emphasize that the phase space is $(\phi_{R},\phi_{I},\pi_{2R},\pi_{2I})$. Applying strongly
the conditions (\ref{conreuno}) and (\ref{conredos}) in the density (\ref{hcanonica}), we obtain the Hamiltonian density
\begin{eqnarray}\label{cr}
 \mathcal{H}_{CKG}=\frac{(m_{1}^{2}-m_{2}^{2})}{2}\pi_{2R}^{2}+\frac{m_{2}^{2}(m_{1}^{2}-m_{2}^{2})}{2}\phi_{R}^{2}+
 \frac{(m_{1}^{2}-m_{2}^{2})}{2}(\triangledown\phi_{R})^{2}
\\\nonumber
+\frac{(m_{1}^{2}-m_{2}^{2})}{2}\pi_{2I}^{2}+\frac{m_{1}^{2}(m_{1}^{2}-m_{2}^{2})}{2}\phi_{I}^{2}+
\frac{(m_{1}^{2}-m_{2}^{2})}{2}(\triangledown\phi_{I})^{2}.
\end{eqnarray}
The Lagrangian density (\ref{a}) with the constraints and reality
conditions is now
\begin{eqnarray}
 \mathcal{L}_{CKG}=(\dot{\phi}\pi_{0}+\dot{\xi}\pi_{2})\lfloor_{cons}-\mathcal{H}_{CKG}
=(m_{1}^{2}-m_{2}^{2})\dot{\phi}_{R}\pi_{2R}\\
+(m_{1}^{2}-m_{2}^{2})\dot{\phi}_{I}\pi_{2I}-\mathcal{H}_{CKG}. \nonumber
\end{eqnarray}
The expression (\ref{cr}) has a direct relationship with the
Hamiltonian density of two real Klein-Gordon fields. The difference
is only a contact transformation, given by
\begin{eqnarray}\label{sigma}
\sigma_{R}=(m_{1}^{2}-m_{2}^{2})^{\frac{1}{2}}\phi_{R}, \qquad  \textmd{\itshape p}_{R}=(m_{1}^{2}-m_{2}^{2})^{\frac{1}{2}}\pi_{2R},
\\ \nonumber
\sigma_{I}=(m_{1}^{2}-m_{2}^{2})^{\frac{1}{2}}\phi_{I}, \qquad
\textmd{\itshape p}_{I}=
(m_{1}^{2}-m_{2}^{2})^{\frac{1}{2}}\pi_{2I}.\nonumber
\end{eqnarray}
Using this transformation in the Hamiltonian density (\ref{cr}) we
get
\begin{eqnarray}\label{kleingordon}
\mathcal{H}_{KG}=\frac{1}{2}\textmd{\itshape p}_{R}^{2}+\frac{m_{2}^{2}}{2}\sigma_{R}^{2}+\frac{1}{2}(\triangledown\sigma_{R})^{2}
+\frac{1}{2}\textmd{\itshape p}_{I}^{2}+\frac{m_{1}^{2}}{2}\sigma_{I}^{2}+\frac{1}{2}(\triangledown\sigma_{I})^{2}.
\end{eqnarray}
The contact transformation (\ref{sigma}) will be very useful in
order to introduce the respective sources in the path integral
formalism.

In next section we will explore in more detail the reality
conditions (\ref{conreuno}) and (\ref{conredos}) since by means of
these structures we will include interaction potentials in this
complex higher order model.

\subsection{Interpretation of the Reality Conditions}

Up to now, we have established that is possible to reduce a complex
higher order system to a first order real system and it results to
be Hermitian at the moment of quantizing. Reality conditions used as
second class constraints \cite{MUV} play a role fundamental in this
reduction  as well as to promote a contact transformation that
allows to recognize the real system as the system of two real
Klein-Gordon fields. However, this method is not the only possible
since our starting point (\ref{conreuno}) was given by hand and not
following a systematic procedure.

Taking this into account we observe that the particular map which
relate the complex phase space $(\phi, \dot{\phi}, \pi_{\phi},
\pi_{\dot{\phi}})$ to the real phase space
$(\psi_{1},\pi_{\psi_{1}},\psi_{2},\pi_{\psi_{2}})$ is
\begin{eqnarray}\label{transcom}
\psi_{1}=\frac{1}{(m_{1}^{2}-m_{2}^{2})^{\frac{1}{2}}}(im_{2}^{2}\phi-i(-\xi+\nabla^{2}\phi)),\nonumber\\
\psi_{2}=\frac{1}{(m_{1}^{2}-m_{2}^{2})^{\frac{1}{2}}}(m_{1}^{2}\phi-(-\xi+\nabla^{2}\phi)),\nonumber\\
\pi_{\psi_{1}}=i\frac{1}{(m_{1}^{2}-m_{2}^{2})^{\frac{1}{2}}}( (\pi_{0}+\nabla^{2}\pi_{2})-m_{1}^{2}\pi_{2}),\nonumber\\
\pi_{\psi_{2}}=\frac{1}{(m_{1}^{2}-m_{2}^{2})^{\frac{1}{2}}}((\pi_{0}+\nabla^{2}\pi_{2})-m_{2}^{2}\pi_{2}.
\end{eqnarray}
In order to show that the phase space $(\psi_{1}, \psi_{2},
\pi_{\psi_{1}}, \pi_{\psi_{2}})$ is real we assume that it is
complex, resulting
\begin{eqnarray}\label{imacero}
(\psi_{1R}+i\psi_{1I})=\frac{1}{(m_{1}^{2}-m_{2}^{2})^{\frac{1}{2}}}[(m_{1}^{2}-m_{2}^{2})\phi_{I}-i\Sigma_{3}],\nonumber\\
(\psi_{2R}+i\psi_{2I})=\frac{1}{(m_{1}^{2}-m_{2}^{2})^{\frac{1}{2}}}[(m_{1}^{2}-m_{2}^{2})\phi_{R}-i\Sigma_{4}],\nonumber\\
(\pi_{\psi_{1R}}+i\pi_{\psi_{1I}})=\frac{1}{(m_{1}^{2}-m_{2}^{2})^{\frac{1}{2}}}[(m_{1}^{2}-m_{2}^{2})\pi_{2I}+i\Sigma_{2}],\nonumber\\
(\pi_{\psi_{2R}}+i\pi_{\psi_{2I}})=\frac{1}{(m_{1}^{2}-m_{2}^{2})^{\frac{1}{2}}}[(m_{1}^{2}-m_{2}^{2})\pi_{2R}+i\Sigma_{1}].
\end{eqnarray}
In conclusion if the phase space is complex, the imaginary part is
proportional to the reality conditions or constraints
(\ref{conreuno}) and (\ref{conredos}) and then the phase space is
real implementing these conditions.

The reality conditions obtained from the complex mapping, that
generates a real phase space, results in reality conditions which
are  no minimal expressions, since
\begin{eqnarray}\label{aconre}
(m_{1}^{2}-m_{2}^{2})\phi^{*}=(m_{1}^{2}+m_{2}^{2})\phi+2(\xi-\nabla^{2}\phi),\nonumber\\
(m_{1}^{2}-m_{2}^{2})\pi_{2}^{*}=-(m_{1}^{2}+m_{2}^{2})\pi_{2}+2(\pi_{0}+\nabla^{2}\pi_{2}),\nonumber\\
(m_{1}^{2}-m_{2}^{2})(\pi_{0}^{*}+\nabla^{2}\pi_{2}^{*})=(m_{1}^{2}+m_{2}^{2})(\pi_{0}+\nabla^{2}\pi_{2})-2m_{1}^{2}
m_{2}^{2}\pi_{2},\nonumber\\
(m_{1}^{2}-m_{2}^{2})(-\xi^{*}+\nabla^{2}\phi^{*})=-(m_{1}^{2}+m_{2}^{2})(-\xi+\nabla^{2}\phi)+2m_{1}^{2}
m_{2}^{2}\phi_{2}
\end{eqnarray}
and by separating in real and imaginary parts
\begin{eqnarray}
\Sigma_{3}+i\Sigma_{4}=0,\qquad \Sigma_{2}+i\Sigma_{1}=0,\nonumber\\
m_{2}^{2}\Sigma_{2}+im_{1}^{2}\Sigma_{4}=0,\qquad  m_{1}^{2}\Sigma_{3}+im_{2}^{2}\Sigma_{4}=0.
\end{eqnarray}
The conditions (\ref{aconre}) can't be used as constraints
\cite{dec} since the conjugated variables in our system aren't
dynamic variables unless we consider the components as real
independent fields in that case (\ref{conreuno}) and
(\ref{conredos}) have a similar complex dynamics to our proposal.

The relationship between the method described with second class constraints or reality conditions and
the method of a complex canonical transformation is
\begin{eqnarray}\label{transosu}
 \sigma_{R}=\psi_{2},\qquad \sigma_{I}=\psi_{1},\nonumber\\
p_{R}=\pi_{2} ,\qquad p_{I}=\pi_{1}.
\end{eqnarray}
Summarizing, starting from the complex Bernard-Duncan model with 12
real degrees of freedom, we reduce it to four real degrees of
freedom. For that we use four constraints that appears from the
definition of the momenta Eq. (\ref{cosmouno}) and four reality
conditions. In this way the reality conditions generate a real first
order theory that is directly related to complex higher order
theory. It is important to highlight that the reality conditions are
second class constraints whose Poisson bracket is a constant matrix.
So at the quantum level we don't have problems to implement these
brackets. This description is incomplete since this model has
neither the self-interactions between the fields nor the interaction
with external fields. In order to introduce these effects in this
model, we will aggregate self-interaction potentials whose
application of the reality conditions result in consistent
potentials into the reduced space.

\section{The Interaction Potentials}
With the purpose of including interactions into the model we take
into account the following criteria: i) The interactions must be
real quantities one's we apply the reality condition and
constraints. ii) In principle it is possible to include real
interactions that dependent of momenta, we don't take into account
this possibility. This possibility modifies the definition of the
momenta and it can result in a non-Lorentz-invariant theory. iii) By
consistency of the theory we require that the interacting
Hamiltonian does not generate new constraints, to implement this
condition we select a set of interactions that commute with the
reality conditions $\Sigma_{1}$ and $\Sigma_{2}$. In this way, it is
possible to choose interaction potentials that are exclusively
dependent of the fields and in addition automatically to commute
with the reality conditions $\Sigma_{3}$ and $\Sigma_{4}$. This
criterion assumes that the time evolution of the constraints is not
modified by the interaction terms.

\subsection{Selection Criteria of the Interaction Potentials}
In order to study these criteria we going to consider an example in
a way that it will be possible to extend to other interaction
potentials. To select this interaction potential we take into
account that applying the reality conditions this potential results
in a real Lorentz invariant term. In order to find such expression,
we consider the conjugate field $\phi^{*}$  but since it is not a
variable of the system, we replace the conjugate expression using
the reality conditions. In this way, the reality conditions allow to
find consistent real interaction potentials at the reduced space.

Now, it is possible to define a bar field inside the extended space that collapses in a conjugate field
inside the reduced space, resulting
\begin{equation}\label{phicon}
\bar{\phi}=\frac{(m_{1}^{2}+m_{2}^{2})}{(m_{1}^{2}-m_{2}^{2})}\phi-
\frac{2}{(m_{1}^{2}-m_{2}^{2})}(-\xi+\nabla^{2}\phi),
\end{equation}
whenever $m_{1}\neq m_{2}$ and different from zero. The expression
of (\ref{phicon}) is  $\bar{\phi}\neq \phi^{*}$ on the extended
space in general, but it can be reduced to the conjugate field
inside the reduced space, resulting
\begin{equation}
\phi^{*}=\bar{\phi}|_{rec}=\phi_{R}-i\phi_{I}.
\end{equation}
This reduced space is gotten of applying the reality conditions
$\Sigma_{3}$ and $\Sigma_{4}$ on the extended space which we will
denote by $|_{rec}$.

The last statement suggests that the selection of interaction
potentials inside the extended space is dependent of the reality
conditions in such a way that by restricting the phase space the
interaction potentials don't leave the reduced space. The components of field in terms of $\phi$ and his conjugate $\bar{\phi}$ are
\begin{eqnarray}\label{conjugado}
 \phi_{R}=\frac{1}{2}(\phi+\bar{\phi})\mid_{cre},\qquad \phi_{I}=\frac{1}{2i}(\phi-\bar{\phi})\mid_{cre}
\end{eqnarray}
that allow us to introduce the possible interaction potentials
\begin{eqnarray}\label{intercab}
  U^{1}_{int}(\phi, \xi)=\int\!d^{3}x\frac{g_{1}}{4!(m_{1}^{2}-m_{2}^{2})^{2}}[m_{1}^{2}\phi+(\xi-\nabla^{2}\phi)]^{4},\\
U^{2}_{int}(\phi, \xi)=\int\!d^{3}x\frac{g_{2}}{4!(m_{1}^{2}-m_{2}^{2})^{2}}[m_{2}^{2}\phi+(\xi-\nabla^{2}\phi)]^{4},\nonumber\\
 U^{3}_{int}(\phi, \xi)=\int\!d^{3}x\frac{g_{3}}{4!(m_{1}^{2}-m_{2}^{2})^{2}}[m_{1}^{2}\phi+(\xi-\nabla^{2}\phi)]^{2}
[m_{2}^{2}\phi+(\xi-\nabla^{2}\phi)]^{2}.\nonumber
\end{eqnarray}
Including these expressions we obtain an interacting Hamiltonian
density that is called total Hamiltonian density and if we commute
it with the reality conditions we obtain proportional elements to
the reality conditions since every potential is real.

The interaction potentials (\ref{intercab}) applying the reality conditions (\ref{conredos})  are
\begin{eqnarray}\label{unoatres}
 U^{1}_{int}\mid_{cre}=\int\!d^{3}x\frac{g_{1}}{4!}\psi_{2}^{4},\qquad U^{2}_{int}\mid_{cre}=\int\!d^{3}x
\frac{g_{2}}{4!}\psi_{1}^{4},\\
U^{3}_{int}\mid_{cre}=\int\!d^{3}x\frac{g_{3}}{4!}\psi_{2}^{2}\psi_{1}^{2}.\nonumber
\end{eqnarray}
The last procedure establishes a way of introducing
self-interactions which can be applied in a systematic form. In the
next section we shall explore another kind of interaction that is
possible to introduce and we will consider the path integral
quantization of the model.

\subsubsection{Some Interaction Potentials}
Following the last procedure, we can consider some different interaction potentials that have the same characteristic in common. From the Hamiltonian formalism we write the real and imaginary parts in terms of the fields (\ref{conjugado}),
resulting
\begin{eqnarray}\label{cuatroasiete}
 U^{4}_{int}(\phi, \xi)=\int\!d^{3}x\frac{-g_{4}}{(m_{1}^{2}-m_{2}^{2})^{\frac{3}{2}}}(m_{1}^{2}\phi+\xi-\nabla^{2}\phi)
(m_{2}^{2}\phi+\xi-\nabla^{2}\phi)^{2},\nonumber\\
U^{5}_{int}(\phi, \xi)=\int\!d^{3}x\frac{g_{5}}{(m_{1}^{2}-m_{2}^{2})^{\frac{3}{2}}}(m_{1}^{2}\phi+\xi-\nabla^{2}\phi)
^{3},\\
U^{6}_{int}(\phi, \xi)=\int\!d^{3}x\frac{-ig_{6}}{(m_{1}^{2}-m_{2}^{2})^{\frac{3}{2}}}(m_{2}^{2}\phi+\xi-\nabla^{2}\phi)
^{3},\nonumber\\
U^{7}_{int}(\phi, \xi)=\int\!d^{3}x\frac{-ig_{7}}{(m_{1}^{2}-m_{2}^{2})^{\frac{3}{2}}}(m_{2}^{2}\phi+\xi-\nabla^{2}\phi)
(m_{1}^{2}\phi+\xi-\nabla^{2}\phi)^{2},\nonumber
\end{eqnarray}
where by applying the reality conditions (\ref{conredos}), we obtain
\begin{eqnarray}
 U^{4}_{int}\mid_{cre}=g_{4}\psi_{2}\psi_{1}^{2},\qquad U^{5}_{int}\mid_{cre}=g_{5}\psi_{2}^{3},\\
U^{6}_{int}\mid_{cre}=g_{6}\psi_{1}^{3},\qquad U^{7}_{int}\mid_{cre}=\psi_{1}\psi_{2}^{2}.\nonumber
\end{eqnarray}
It is important to emphasize that the reality conditions or second
class constraints are a closed set under the time evolution. We
showed it in the equations (\ref{imacero}) where was exhibited that
the imaginary part of the fields $\psi$'s vanish module the reality
conditions (\ref{conreuno}), (\ref{conredos}). Now, if we include
the potentials (\ref{unoatres}) and (\ref{cuatroasiete}) to the
Hamiltonian density (\ref{hcanonica}), the evolution of the reality
conditions is not modified by interactions. This will be helpful in
order to build the path integral in such a way that the sources are
consistent with the theory. Furthermore, considering the dependence
between the fields due to the constraints and reality conditions we
will have to take into account the dependence between the sources.

\section{Path Integral and Complex Sources}
In order to establish the path integral and to introduce the
appropriate sources of the fields, we have to analyze the classical
properties of these sources. These properties express a relationship
between the sources in such way that the imaginary terms are
canceled in the Hamiltonian density and in this form the
Cauchy-Riemann equations will be preserved. With these principles,
the path integral formulation with currents is quite manageable.

\subsection{Equations of Motion with currents}
Let us consider a free complex Hamiltonian density which include the
currents of fields and momenta
\begin{equation}\label{hsfuentes}
 \mathcal{H}_{S}=\mathcal{H}-J\phi-K\pi_{0}-L\eta-M\pi_{1}-N\xi-O\pi_{2}.
\end{equation}
Note that we have included every kind of currents that belong to the
complex Bernard-Duncan model. Using the Cauchy-Riemann conditions
for the complex Hamiltonian density (\ref{hsfuentes}), together with
(\ref{ostrogradsky}),  we obtain the equations of motion
\begin{eqnarray}\label{fuenteem}
 \dot{\phi}_{R}=\frac{\partial \mathcal{H}_{SR}}{\partial\pi_{0R}}=2\pi_{2R}-\eta_{R}-K_{R},
 \qquad\dot{\phi}_{I}=-\frac{\partial \mathcal{H}_{SR}}{\partial\pi_{0I}}=2\pi_{2I}-\eta_{I}-K_{I},\nonumber\\
\dot{\pi}_{0R}=\frac{\partial \mathcal{H}_{SR}}{\partial\phi_{R}}=\nabla^{2}(\nabla^{2}\phi_{R})-(m_{1}^{2}+m_{2}^{2})
\nabla^{2}\phi_{R}+m_{1}^{2}m_{2}^{2}\phi_{R}-J_{R},\\
\dot{\pi}_{0I}=\frac{\partial \mathcal{H}_{SR}}{\partial\phi_{I}}=-\nabla^{2}(\nabla^{2}\phi_{I})+(m_{1}^{2}
+m_{2}^{2})\nabla^{2}\phi_{I}-m_{1}^{2}m_{2}^{2}\phi_{I}+J_{I},\nonumber\\
 \dot{\eta}_{R}=\frac{\partial \mathcal{H}_{SR}}{\partial\pi_{1R}}=\xi_{R}-M_{R},\qquad
 \dot{\eta}_{I}=-\frac{\partial \mathcal{H}_{SR}}{\partial\pi_{1I}}=\xi_{I}-M_{I},\nonumber\\
\dot{\pi}_{1R}=\frac{\partial \mathcal{H}_{SR}}{\partial\eta_{R}}=2\nabla^{2}\pi_{2R}-(m_{1}^{2}+m_{2}^{2})\pi_{2R}+
(m_{1}^{2}+m_{2}^{2})\eta_{R}\nonumber\\
-2\nabla^{2}\eta_{R}-\pi_{0R}-L_{R},\nonumber\\
\dot{\pi}_{1I}=\frac{\partial \mathcal{H}_{SR}}{\partial\eta_{I}}=-2\nabla^{2}\pi_{2I}+(m_{1}^{2}+m_{2}^{2})\pi_{2I}-
(m_{1}^{2}+m_{2}^{2})\eta_{I}\nonumber\\
+2\nabla^{2}\eta_{I}+\pi_{0I}+L_{I},\nonumber\\
\dot{\xi}_{R}=\frac{\partial\mathcal{H}_{SR}}{\partial\pi_{2R}}=2\pi_{0R}+2\nabla^{2}\eta_{R}-(m_{1}^{2}+m_{2}^{2})
\eta_{R}-O_{R},\nonumber\\
\dot{\xi}_{I}=-\frac{\partial\mathcal{H}_{SR}}{\partial\pi_{2I}}=2\pi_{0I}+2\nabla^{2}\eta_{I}-(m_{1}^{2}
+m_{2}^{2})\eta_{I}-O_{I},\nonumber\\
\dot{\pi}_{2R}=\frac{\partial\mathcal{H}_{SR}}{\partial\xi_{R}}=\pi_{1R}-\xi_{R}-N_{R},\qquad
\dot{\pi}_{2I}
=\frac{\partial\mathcal{H}_{SR}}{\partial\xi_{I}}=-\pi_{1I}+\xi_{I}+N_{I}.\nonumber
\end{eqnarray}
Using the equations of motion (\ref{fuenteem}), the four constraints
resulting of the momenta (\ref{cosmouno}) and, the four reality
conditions (\ref{conreuno}) and (\ref{conredos}), we obtain that the
currents are related by
\begin{eqnarray}\label{fuenteuno}
 J_{R}=(-\nabla^{2}+m_{1}^{2})N_{R},\qquad J_{I}=(-\nabla^{2}+m_{2}^{2})N_{I}\\
L_{R}+O_{R}=(\nabla^{2}-m_{2}^{2})K_{R},\qquad L_{I}+O_{I}=(\nabla^{2}-m_{1}^{2})K_{I}.\nonumber
\end{eqnarray}
From the expression (\ref{hsfuentes}) the currents are independent
quantities into the extended space, but if we consider the reduced
space defined by the constraints and the reality conditions. The
currents aren't independent and we obtain relationships between them
(\ref{fuenteuno}). In our procedure, the relationships are
established by the constraints and reality conditions and the total
derivative plays a fundamental role in order to define the fields
and the currents.

\subsection{Path Integral}
In this section, we will be concerned with establishing the complex
higher order theory in terms of the path integral formalism, in
order to get a consistent quantization that includes interaction
potentials. To introduce the path integral, we shall use an
integration measure that considers every constraint and reality
condition. Taking as starting point the complex Hamiltonian density
with complex currents (\ref{hsfuentes}), we will build the
integration measure on the path integral following the Senjanovic's
method \cite{Senja}. However, it is necessary to add independently
the classical relations between the currents, since they are
classical fields and are not quantized in the usual description. It
is important to mention that the source terms into the Hamiltonian
density are complex quantities that obey the equations of
Cauchy-Riemann for multiple variables, that are handled in the
expressions above (\ref{fuenteuno}), in such a way that the
imaginary part of the Hamiltonian density is zero including every
constraint and reality condition.

To begin with, we study the path integral with currents that describe the annihilation and creation process.
In order to include the currents, we choose the respective complex fields, $\phi$, $\eta$, $\xi$ and their
respective momenta $\pi_{0}$, $\pi_{1}$, $\pi_{2}$. So, to introduce the measure of integration,  we consider in
the path integral that  the real and imaginary parts are independent, and we include as functional Dirac's deltas every second class constraint resulting of the momenta (\ref{cosmouno})
together with the reality conditions (\ref{conreuno}) and
(\ref{conredos}).

To apply the Senjanovic's method we define the set of constraints as
\begin{equation}
\Omega_{i}=(\gamma_{1}, \gamma_{2}, \gamma_{3}, \gamma_{4}, \Sigma_{1}, \Sigma_{2}, \Sigma_{3}, \Sigma_{4})
\end{equation}
that allows to establish the respective determinant of the Poisson brackets, resulting
\begin{equation}
 \{ \Omega_{i} ,\Omega_{j} \}= \mathcal{G}_{ij}\delta(\vec{x}-\vec{x'}),
\end{equation}
with
\begin{equation}\label{detp}
\det \left( \mathcal{G}_{ij}\right) =(m_{1}^{2}-m_{2}^{2})^{4}.
\end{equation}
The integration measure on the path integral is
\begin{equation}\label{muuno}
 \mathcal{D}\mu=\mathcal{D}\Theta^{a}_{R}\mathcal{D}\Pi_{aR}\mathcal{D}\Theta^{a}_{I}\mathcal{D}\Pi_{aI}
\det\mid\left\lbrace\Omega_{i},\Omega_{j}\right\rbrace\mid \prod_{i}^{8}
\delta(\Omega_{i})\\
\end{equation}
where $\prod_{i}$ regards a product of Dirac deltas on each constraint.

The currents in terms of this notation are
\begin{eqnarray}
\mathcal{J}_{aA}=(J_{R}, L_{R}, N_{R}, J_{I}, L_{I}, N_{I})\\
\mathcal{K}^{b}_{B}=(K_{R}, M_{R}, O_{R}, K_{I}, M_{I}, O_{I})\nonumber
\end{eqnarray}
with $a =J, L, N$, $b =K, M, O$ and $A,B =R,I$. The generating
functional using the above elements is
\begin{eqnarray}\label{path}
Z=\int\mathcal{D}\mu
\exp [i\int\!d^{4}x (\dot{\Theta}^{a}_{R}\Pi_{aR}-\dot{\Theta}^{a}_{I}\Pi_{aI}
-\mathcal{H}_{R}
+\Theta^{a}_{R}\mathcal{J}_{aR}-\Theta^{a}_{I}\mathcal{J}_{aI}\\
\nonumber
+\Pi_{aR}\mathcal{K}^{a}_{R}-\Pi_{aI}\mathcal{K}^{a}_{I})+i(\dot{\Theta}^{a}_{R}\Pi_{aI}+\dot{\Theta}^{a}_{I}\Pi_{aR}-\mathcal{H}_{I}+\Theta^{a}_{R}\mathcal{J}_{aI}+\Theta^{a}_{I}\mathcal{J}_{aR}\\\nonumber
+\Pi_{aR}\mathcal{K}^{a}_{I}+\Pi_{aI}\mathcal{K}^{a}_{R})],\nonumber
\end{eqnarray}
where $\mathcal{H}_{R}$ and $\mathcal{H}_{I}$ are given in
(\ref{hamrealim}) and (\ref{hamrealimdos}) respectively and we use
the notation given in (\ref{aaa}), (\ref{bbb}). We must notice that
the determinant of the second class constraints only contribute to
the generating functional with a constant factor proportional to the
difference of the masses of the scalar fields. This factor is not
important in the case of different masses, but perhaps is quite
relevant in the limit of equal masses.

Using the the Dirac's delta functional in (\ref{path}) we integrate
over the fields \break
$(\eta_R,\eta_I,\xi_R,\xi_I,\pi_{0R},\pi_{0I},\pi_{1R},\pi_{1I})$.
From this the path integral is reduced to
\begin{eqnarray}\label{intecamtres}
Z_{R}=\int\mathcal{D}\phi_{R}\mathcal{D}\phi_{I}\mathcal{D}\pi_{2R}\mathcal{D}\pi_{2I} \exp\{i\int\!d^{4}x\left(
[(m_{1}^{2}-m_{2}^{2})\dot{\phi}_{R}\pi_{2R}+(m_{1}^{2}-m_{2}^{2})
\right.\nonumber\\
\left.\dot{\phi}_{I}\pi_{2I} -\mathcal{H}_{CKG}+\phi_{R}(\nabla^{2}N_{R}-m_{2}^{2}N_{R}+J_{R})+\phi_{I}
(-\nabla^{2}N_{I}+m_{1}^{2}N_{I}-J_{I}) \right. \nonumber\\
\left. +\pi_{2R}(-\nabla^{2}K_{R}+m_{1}^{2}K_{R}+L_{R}+O_{R})+\pi_{2I}(\nabla^{2}K_{I}-m_{2}^{2}K_{I}-L_{I}-O_{I})]
\right. \nonumber\\
\left.+i[\phi_{R}(\nabla^{2}N_{I}-m_{2}^{2}N_{I}+J_{I})
+\phi_{I}(\nabla^{2}N_{R}-m_{1}^{2}N_{R}+J_{R})\right.\nonumber\\
\left.+\pi_{2I}(-\nabla^{2}K_{R}+m_{2}^{2}K_{R}+L_{R}+O_{R})\right.\nonumber\\
\left.+\pi_{2R}(-\nabla^{2}K_{I}+m_{1}^{2}K_{I}+L_{I}+O_{I})]\right) \}.
\end{eqnarray}
and since the Hamiltonian density is still complex, this expression
does not include completely the final map to the real space.

Up to this point, we have the path integral without constraints. The
benefit of preserving the imaginary part of the Hamiltonian density,
is that in the imaginary part of (\ref{intecamtres}), we obtain the
equations (\ref{fuenteuno}) as a consequence of the measure of the
path integral. In conclusion, the imaginary part, that could generate
ghosts, disappears from the path integral using the classical
relationship of the currents (\ref{fuenteuno}). In this way, we obtain
\begin{eqnarray}\label{pathin}
Z_{R}=\int\mathcal{D}\phi_{R}\mathcal{D}\phi_{I}\mathcal{D}\pi_{2R}\mathcal{D}\pi_{2I} \exp\{i\int\!d^{4}x
\left( [(m_{1}^{2}-m_{2}^{2})\dot{\phi}_{R}\pi_{2R}\right.\\\nonumber
\left.+(m_{1}^{2}-m_{2}^{2})
\dot{\phi}_{I}\pi_{2I} -\mathcal{H}_{CKG}+(m_{1}^{2}-m_{2}^{2})\phi_{R}N_{R}+(m_{1}^{2}-m_{2}^{2})\phi_{I}N_{I}\right.
\\\nonumber
\left.+(m_{1}^{2}-m_{2}^{2})\pi_{2R}K_{R}+(m_{1}^{2}-m_{2}^{2})\pi_{2I}K_{I}]\right) \}.\nonumber
\end{eqnarray}
Finally, we can apply  the contact transformation
\begin{eqnarray}
\sigma_{R}=(m_{1}^{2}-m_{2}^{2})^{\frac{1}{2}}\phi_{R}, \qquad \textmd{\itshape p}_{R}=(m_{1}^{2}-m_{2}^{2})^{\frac{1}{2}}\pi_{2R},
\\ \nonumber
\sigma_{I}=(m_{1}^{2}-m_{2}^{2})^{\frac{1}{2}}\phi_{I}, \qquad \textmd{\itshape p}_{I}=(m_{1}^{2}-m_{2}^{2})^{\frac{1}{2}}\pi_{2I},
\\ \nonumber
\aleph_{R}=(m_{1}^{2}-m_{2}^{2})^{\frac{1}{2}}N_{R}, \qquad \aleph_{I}=(m_{1}^{2}-m_{2}^{2})^{\frac{1}{2}}N_{I},
\\ \nonumber
\Bbbk_{R}=(m_{1}^{2}-m_{2}^{2})^{\frac{1}{2}}K_{R}, \qquad \Bbbk_{I}=(m_{1}^{2}-m_{2}^{2})^{\frac{1}{2}}K_{I},
\nonumber
\end{eqnarray}
and the Jacobian of this transformation is
\begin{eqnarray}\label{Jaco}
\mid J\mid=\frac{1}{(m_{1}^{2}-m_{2}^{2})^{2}}, \qquad
 d\phi_{R}d\pi_{2R} d\phi_{I}d\pi_{2I}=\mid J\mid d\sigma_{R} d\textmd{\itshape p}_{R}d\sigma_{I} d\textmd{\itshape p}_{I}.
\end{eqnarray}
The expressions (\ref{Jaco}) transform the path integral (\ref{pathin}) into
\begin{eqnarray}\label{zkg}
Z_{KG}=S\int\mathcal{D}\sigma_{R}\mathcal{D}\sigma_{I}\mathcal{D}\textmd{\itshape p}_{R}\mathcal{D}\textmd{\itshape p}_{I} \exp\{i\int\!d^{4}x
\left(\dot{\sigma}_{R}\textmd{\itshape p}_{R}
\left.+\dot{\sigma}_{I}\textmd{\itshape p}_{I} \right.\right.
\\\nonumber
\left.-\mathcal{H}_{KG}(\sigma_{R},\sigma_{I},\textmd{\itshape p}_{R},\textmd{\itshape p}_{I})+\sigma_{R}\aleph_{R}+\sigma_{I}\aleph_{I}
+\textmd{\itshape p}_{R}\Bbbk_{R}+\textmd{\itshape p}_{I}\Bbbk_{I})\right \}\nonumber
\end{eqnarray}
where $\mathcal{H}_{KG}$ is the expression (\ref{kleingordon}). In
expression (\ref{zkg}),  $S$ is a constant quantity  that  results
from the determinant (\ref{detp}) and the jacobian expression
(\ref{Jaco}) resulting from the contact transformation. It will
allow to define the generating functional of interactions using the
free generating functional.

\subsubsection{Generating Functional of Interactions}
Using the free generating functional (\ref{zkg}) it is possible to
build the generating functional with interactions and we obtain

\begin{eqnarray}
Z_{int}=\int\mathcal{D}\sigma_{R}\mathcal{D}\sigma_{I}\mathcal{D}\textmd{\itshape p}_{R}\mathcal{D}\textmd{\itshape p}_{I} \exp\{i\int\!d^{4}x\left(
\dot{\sigma}_{R}\textmd{\itshape p}_{R}+\dot{\sigma}_{I}\textmd{\itshape p}_{I}\right.\\\nonumber
\left. -\mathcal{H}_{CKG}-\frac{g_{1}}{4!}\sigma_{R}^{4}-\frac{g_{2}}{4!}\sigma_{I}^{4}
-\frac{g_{3}}{4!}\sigma_{R}^{2}\sigma_{I}^{2}
+\sigma_{R}\aleph_{R}
\right.\\\nonumber
\left.+\sigma_{I}\aleph_{I}
+\textmd{\itshape p}_{R}\Bbbk_{R}+\textmd{\itshape p}_{I}\Bbbk_{I}\right) \}\nonumber
\end{eqnarray}
that in terms of the free generating functional are
\begin{eqnarray}
Z_{int1}=exp\{-\frac{ig_{1}}{4!}\int\!d^{4}x(\frac{\delta}{i\delta\aleph_{R}})^{4}\}Z_{KG},\\
Z_{int2}=exp\{-\frac{ig_{2}}{4!}\int\!d^{4}x(\frac{\delta}{i\delta\aleph_{I}})^{4}\}Z_{KG},\\
Z_{int3}=exp\{-\frac{ig_{3}}{4!}\int\!d^{4}x(\frac{\delta}{i\delta\aleph_{R}})^{2}
(\frac{\delta}{i\delta\aleph_{I}})^{2}\}Z_{KG},
\end{eqnarray}
and it shows that is possible to introduce interactions in the high
order derivative theory whereas these interactions are compatible
with the constraints and the reality conditions. In this way, our
procedure allows to quantize the complex higher order theory
(\ref{a}), with the interactions (\ref{intercab}). In next section,
we will apply this method to the Schwinger model in order to check
for an explicit example how it works.

\section{Using the Method}\label{ejemplos}
In order to explore the scope of the method that includes the
boundary conditions  $(\phi, \ddot{\phi})$, we analyze a concrete
example that shows limit cases of our description. Between all the  possible
examples, we have the Schwinger model that is a good starting point since it has been explored exhaustively and is a very
important model, if we want to study higher order time derivative
theories. In this electrodynamics into two dimensions, we find a
phenomenon known as Bosonization that is easily viewed using the
higher order time derivative theories \cite{schwinger}. In this
section,  we apply the method here described, starting from the general description
of the 2-dimensional electrodynamics to a real higher order theory
that is a particular case of our complex model.

\subsection{The Schwinger model}
In this section, we apply the above method to the Schwinger model,
and we compare with a previous  procedure \cite{schwinger}.

Consider  the Schwinger model that is a formulation of the massless
electrodynamics in $1+1$ dimensions. The Lagrangian density of
departure is
\begin{equation}\label{schwingeruno}
\mathcal{L}_{ED}=-\frac{1}{4}F_{\mu\nu}F^{\mu\nu}+\bar{\psi}[i\gamma^{\mu}(\partial_{\mu}-ieA_{\mu})]\psi,
\end{equation}
with $F_{\mu\nu}=\partial_{\mu}A_{\nu}-\partial_{\nu}A_{\mu}$. This
electromagnetic Lagrangian density is coupled to a Dirac Lagrangian
density with zero mass. On the classical level it is well known that
the vector current is conserved, but we must also consider the
chiral current that results classically to be also a conserved
quantity, since the electron has zero mass. But the focus is on the
respective quantized model. In this case the chiral current is not
more a conserved quantity resulting in a breaking of the classical
symmetry. This phenomenon is known as chiral anomaly and its
consequences can be studied by means of a non local transformation
of (\ref{schwingeruno}) that is given by
\begin{eqnarray}\label{lore}
A_{\mu}=-\frac{1}{e}\varepsilon_{\mu\nu}\partial^{\nu}\varphi+\frac{1}{e}\partial_{\mu}\eta,
\qquad \psi=\exp(i\gamma^{5}\varphi+i\eta)\kappa,\\
\bar{\psi}=\bar{\kappa}\exp(i\gamma^{5}\varphi-i\eta).\nonumber
\end{eqnarray}
with two-dimensional Dirac matrices
\begin{eqnarray}
 [\gamma^{\mu},\gamma^{\nu}]=2g^{\mu\nu}, \qquad \gamma^{\mu}\gamma^{5}=\epsilon^{\mu\nu}\gamma_{\nu},\nonumber\\
\epsilon_{01}=-\epsilon_{10}.
\end{eqnarray}
Using the expressions (\ref{lore}) we get
\begin{equation}
 F_{\mu\nu}=\frac{1}{e}\varepsilon_{\mu\nu}\Box\varphi, \qquad \Box \eta=0.
\end{equation}
and the gauge transformation is expressed in terms of the field
$\eta$
\begin{equation}
A'_{\mu}(x)=A_{\mu}(x)+\partial_{\mu}\Lambda(x)=-\frac{1}{e}\varepsilon_{\mu\nu}\partial^{\nu}\phi+\frac{1}{e}\partial_{\mu}(\eta+e\Lambda)=-\frac{1}{e}\varepsilon_{\mu\nu}\partial^{\nu}\phi+\frac{1}{e}\partial_{\mu}\eta',
\end{equation}
implying that this field doesn't appears in the Lagrangian density.
The classical Lagrangian density using the transformation
(\ref{lore})  is
\begin{equation}
 \mathcal{L}=\frac{1}{2e^{2}}(\Box\varphi)^{2}+\bar{\kappa}i\gamma^{\mu}\partial_{\mu}\kappa,
\end{equation}
where it has been decouple the bosonic part given by the field
$\varphi$ from the fermionic part given by  the field $\kappa$. In
order to realize the quantization, it is used  the path integral in
which the non local transformation is applied (\ref{lore}), implying
the appearance of the chiral anomaly in the  associated
Faddeev-Popov Jacobian. To cancel this anomaly the total Lagrangian density acquires a new
term. Our starting point is precisely this full effective action
giving place to
\begin{equation}
S_{0}=\int \! d^{4}x\frac{1}{2}[-(\Box\varphi)^{2}+m_{1}^{2}\partial_{\mu}\varphi\partial^{\mu}\varphi]
\end{equation}
with the field $\varphi$ real. An important fact to bear in mind is
that the Schwinger model is a particular case of the Bernard-Duncan
model  (\ref{s}) with  $m^{2}_{2}=0$ and
$m_{1}^{2}=\frac{e^{2}}{\pi}$. The next step is to realize the
complexification as we described previously including the total
derivative, with the result
\begin{equation}
S=\int \! d^{4}x\frac{1}{2}[(\Box\phi)^{2}+m_{1}^{2}\partial_{\mu}\phi\partial^{\mu}\phi]+
\partial_{\mu}\phi(\partial^{\mu}\Box\phi)
\end{equation}
and applying a variation of the action in order to obtain the
momenta, we have
\begin{equation}
 \pi_{0}=\phi^{(3)}-2\nabla^{2}\dot{\phi}+m_{1}^{2}\dot{\phi},\qquad\pi_{1}=0,\qquad
\pi_{2}=\dot{\phi}.
\end{equation}
With these momenta we can obtain the respective  Hamiltonian density
\begin{eqnarray}
 \mathcal{H}_{S}=\pi_{1}\xi+\pi_{0}\pi_{2}+2\pi_{2}\nabla^{2}\eta-m_{1}^{2}\eta\pi_{2}
-\frac{1}{2}\xi^{2}-\frac{1}{2}(\nabla^{2}\phi)^{2}+\xi\nabla^{2}\phi\\\nonumber
-\frac{m_{1}^{2}}{2}\eta^{2}+\frac{m_{1}^{2}}{2}\nabla\phi\cdotp\nabla\phi
-\eta\nabla^{2}\eta+m_{1}^{2}\eta^{2}-\xi\nabla^{2}\phi+(\nabla^{2}\phi)^{2}\\ \nonumber
+\pi_{0}\pi_{2}-\eta\pi_{0}
\end{eqnarray}
and, we obtain the respective constraints
\begin{eqnarray}
 \gamma_{1}=\pi_{1R}, \qquad \gamma_{2}=\pi_{1I},\\
\gamma_{3}=\pi_{2R}-\eta_{R},\qquad
\gamma_{4}=\pi_{2I}-\eta_{I}.\nonumber
\end{eqnarray}
These restrictions are second class constraints. Using these
constraints we define the respective Dirac brackets
\begin{eqnarray}
 \{ \phi_{R}(t,\vec{x}),\pi_{0R}(t,\vec{x}_{0})\}^{*}=\delta^{3}(\vec{x}-\vec{x}_{0})\\
\nonumber
 \{ \phi_{I}(t,\vec{x}),\pi_{0I}(t,\vec{x}_{0})\}^{*}=-\delta^{3}(\vec{x}-\vec{x}_{0})\\
\nonumber
\{ \eta_{R}(t,\vec{x}),\xi_{R}(t,\vec{x}_{0})\}^{*}=-\delta^{3}(\vec{x}-\vec{x}_{0})\\\nonumber
 \{ \eta_{I}(t,\vec{x}),\xi_{I}(t,\vec{x}_{0})\}^{*}=\delta^{3}(\vec{x}-\vec{x}_{0})\\
\nonumber
\{ \xi_{R}(t,\vec{x}),\pi_{2R}(t,\vec{x}_{0})\}^{*}=\delta^{3}(\vec{x}-\vec{x}_{0})\\
\nonumber
 \{ \xi_{I}(t,\vec{x}),\pi_{I}(t,\vec{x}_{0})\}^{*}=-\delta^{3}(\vec{x}-\vec{x}_{0}).\\
\nonumber
\end{eqnarray}
The second class constraints are applied strongly to the Hamiltonian
density ones the Dirac brackets are imposed. The resulting
Hamiltonian density is
\begin{equation}\label{hams}
\mathcal{H}_{C}=\pi_{0}\pi_{2}-\frac{1}{2}\xi^{2}-\frac{m_{1}^{2}}{2}\pi_{2}^{2}
+\frac{1}{2}(\nabla^{2}\phi)^{2}
+\frac{m_{1}^{2}}{2}(\nabla\phi)^{2}+\pi_{2}\nabla^{2}\pi_{2}.
\end{equation}
The following step is to impose the reality conditions in order to
reduce the complex space to real space and check if the time
evolution does not generate another constraint. In this case, the
complete set of reality conditions is
\begin{eqnarray}
 \Sigma_{1}=\pi_{0I}+\nabla^{2}\pi_{2I},\qquad \Sigma_{4}=-\xi_{I}+\nabla^{2}\phi_{I}-m_{1}^{2}\phi_{I}\\
\Sigma_{2}=\pi_{0R}+\nabla^{2}\pi_{2R}-m_{1}^{2}\pi_{2R},\qquad \Sigma_{3}=-\xi_{R}+\nabla^{2}\phi_{R}\nonumber
\end{eqnarray}
where $\Sigma_{1}$ with $\Sigma_{2}$ are two arbitrary constraints and $\Sigma_{3}$ with
$\Sigma_{4}$ are consequence of the time evolution of the first two constraints.
These second class constraints imply the next Dirac brackets
\begin{eqnarray}
\left\lbrace  \phi_{R}(t,\vec{x}),\pi_{2R}(t,\vec{x}_{0})  \right\rbrace^{**}=\frac{1}{m_{1}^{2}}
\delta^{3}(\vec{x}-\vec{x}_{0}),\\\nonumber
\left\lbrace  \phi_{I}(t,\vec{x}),\pi_{2I}(t,\vec{x}_{0}) \right\rbrace^{**}=\frac{1}{m_{1}^{2}}
\delta^{3}(\vec{x}-\vec{x}_{0}).\nonumber
\end{eqnarray}
The real Hamiltonian density is
\begin{eqnarray}
 \mathcal{H}_{CKG}=\frac{m_{1}^{2}}{2}\pi_{2I}^{2}+\frac{m_{1}^{2}}{2}(\triangledown\phi_{I})^{2}
+\frac{m_{1}^{4}}{2}\phi_{I}^{2}
+\frac{m_{1}^{2}}{2}\pi_{2R}^{2}+\frac{m_{1}^{2}}{2}(\triangledown\phi_{R})^{2}
\end{eqnarray}
This Hamiltonian density isn't a higher order time derivative theory
and is a real quantity. Now, it is possible to define a contact
transformation
\begin{equation}
\sigma_{R}=m_{1}\phi_{R}, \qquad \textmd{\itshape
p}_{R}=m_{1}\pi_{2R},\qquad \sigma_{I}=m_{1}\phi_{I}, \qquad
\textmd{\itshape p}_{I}=m_{1}\pi_{2I}.
\end{equation}
Using the above expressions the new  Dirac brackets are
\begin{eqnarray}
\left\lbrace  \sigma_{R}(t,\vec{x}),\textmd{\itshape p}_{R}(t,\vec{x}_{0})  \right\rbrace^{**}=\delta^{3}(\vec{x}-\vec{x}_{0}),\\
\nonumber
\left\lbrace  \sigma_{I}(t,\vec{x}),\textmd{\itshape p}_{I}(t,\vec{x}_{0}) \right\rbrace^{**}=\delta^{3}(\vec{x}-\vec{x}_{0}),
\nonumber
\end{eqnarray}
and the Hamiltonian density is
\begin{eqnarray}\label{kgs}
 \mathcal{H}_{KG}=\frac{1}{2}\textmd{\itshape p}_{I}^{2}+\frac{1}{2}(\triangledown\sigma_{I})^{2}
+\frac{m_{1}^{2}}{2}\sigma_{I}^{2}
+\frac{1}{2}\textmd{\itshape p}_{R}^{2}+\frac{1}{2}(\triangledown\sigma_{R})^{2}.
\end{eqnarray}
In this way, we see the relationship between our real Hamiltonian
density (\ref{kgs}) with our complex higher order time derivative
Hamiltonian density (\ref{hams}). This relationship shows the
complex Schwinger model with the border conditions
$(\phi,\ddot{\phi})$ is linked to the Hamiltonian density of two
real Klein-Gordon fields, a massless field  and other massive
excitation.  This result is well known \cite{Swieca} and in this way
our procedure reproduces the correct result. Our method is also
applicable to another boundary condition in such a way that it is
possible to fix $(\phi,\dot{\phi})$ without the total derivative.

\section{Conclusions}
In this work, it was introduced a method that makes it possible to
map from a complex higher order derivative theory with interaction
potentials, the complex interacting Bernard-Duncan model, to an
interacting theory of two real Klein-Gordon fields. This complex
extension to the higher order derivative theory is a consequence of
thinking in a more general theory with a complex structure that is
possible to quantize using the reality conditions and avoiding
problems as unbounded energy and states of negative norm. The
complex theory allows to gain more flexibility in order to establish
a different concept of hermiticity, by means of reality condition,
resulting of a complex classical mechanics to describe a higher
order derivative theory. The key point was to establish a complex
structure into the Hamiltonian formalism using Ostrogradsky method
and to develop a mapping mechanism to a real space. The basic idea
is to use the reality conditions \cite{ashtekar}, corresponding to
this complex theory, as second class constraint and to aggregate
interaction potentials that are mapped to real quantities applying
these conditions. It is not easy to set the interaction potentials
because they must not generate additional conditions that constrain
the degrees of freedom when the temporal evolution is done. Moreover
we showed that is possible to set up a complex description using the
classical mechanics, stated by Ostrogradsky, consistent with
Hamilton's equations and Cauchy-Riemann equations. The information
that is given by this complex description is predetermined in such a
way that the complex structure of multi variables is respected and
in consequence the Cauchy-Riemann equations are satisfied. Another
step in the method was to raise the derivative order of the theory
in such way to select the appropriated boundary conditions. The
original order of the theory is recovered through second class
constraints using the Dirac's theory of constraints
 \cite{Dir, Hen}. However, this complex higher order derivative theory
has double degrees of freedom when compared to the real
Bernard-Duncan theory. In order to reduce this extra degrees of
freedom, we directly introduce two constraints that evolve in
another two constraints resulting in a complete set of second class
that reduces this theory to a real description. These constraints
are the reality conditions.

From this description, the quantization is possible using the path
integral with the Senjanovic's method \cite{Senja} developed to
quantize a theory with second class constraints. Using complex
currents and developing a formalism that includes them separating in
components the constraints are applied and the relationship between
higher order fields is exhibited resulting from a complex structure
and reality conditions. The complex currents are not independent in
this higher order derivative model if reality conditions are
established. This is consistently with Ostrodrasky's method
\cite{os}. As a final point on the path integral applied to the
complex Bernard-Duncan model with currents, we conclude that it can
reduce to quantize two Klein-Gordon fields with currents using the
path integral and applying constraints and taking into account a
contact transformation. Furthermore, the cancelation of the
imaginary part of the path integral generates the relations between
the currents which are also obtained classically by means of
Hamiltonian equations of motion including currents.

To include interaction potentials, we considered quantities in the
Hamiltonian density that do not generate new constraints and they
must preserve a closed set of constraints when the temporal
evolution is established (\ref{intercab}). These interaction
potentials are generated by currents either in the complex space or
real space using the reality conditions. The interaction potentials
be attached to the Hamiltonian density result in a renormalizable
theory that is defined into the reduced space (\ref{unoatres})
\cite{Margalli}.Finally, we used that method in the Schwinger model
\cite{schwinger} obtaining consistent results.

\end{document}